# Sum-Rate Performance of Millimeter Wave MIMO Shared Spectrum Systems

Abhishek Agrahari and Aditya K. Jagannatham *Member, IEEE*

*Abstract*—This paper study the effect of interference management arising due to millimeter wave (mmWave) communication on the performance of multiple input multiple output (MIMO) based spectrum sharing cognitive systems. The highly directed mmWave signals have short wavelength property which enables to fabricate an increased number of antennas in a small confined space of the communication system. The large antenna elements are utilized to harvest high beamforming gains and to improve the signal to noise ratio (SNR) by analog processing at the cognitive base station (CBS) and secondary users. Analog processing at the CBS and at secondary user have hardware and computation restrictions. These restrictions are solved by designing radio frequency (RF) precoding and RF combining. In the digital domain, it is shown that interference between secondary users can be canceled by baseband block diagonalization technique. Towards this end, the underlay mode of operation is considered to control interference on the primary user and to derive the expression for optimal power allocation. The derived power allocation is a solution to the rate maximization problem of mmWave MIMO cognitive radio (CR) systems. Finally, numerical simulations are performed to yield various significant insights into for different mmWave CR system conditions and also validate the improvement in performance in comparison to the only digital communication systems.

## I. INTRODUCTION

In recent years, millimeter wave (mmWave) communication has drawn remarkable attention due to its ability to achieve high capacity demands of the fifth generation wireless networks [1], [2], [3]. Moreover, highly directed nature of transmission and sensitivity towards obstacles are the important characteristics of mmWave communication [1], [2]. These characteristics reduces the extent of interference between two different wireless networks in a frequency band of mmWave spectrum [4], [5], [3]. Thus, the portion of mmWave spectrum licensed to primary users can be exploited opportunistically by unlicensed secondary users or cognitive users. In this work, we consider underlay operation mode to maximize the rate performance of mmWave channel based cognitive radio (CR) system. In this underlay CR mode, secondary user operates in the shared mmWave spectrum with primary user as long as the resulting interference at primary user is kept below a certain level [6], [7].

Multiple input multiple output (MIMO) technology is a key towards manifestation of high data rate in the mmWave wireless communication and therefore significant precoding schemes are of fundamental importance because of its potential to harvest antenna gain from the large number of antenna elements [3], [8]. Moreover, when number of antennas are high, the conventional way of precoding is inapplicable in designing a dedicated digital and analog hardware for each antenna element due to the high cost of the hardware and extremely high power consumption [3], [4], [9]. However, this practical limitations can be appropriately encountered by separate analog and digital processing and implementing phase shifters in analog circuity of the communication system.

Analog precoder is formed by a set of beamformers and each of them is transmitting a single data stream in RF domain. The analog RF beamformer steers the beam in the direction of the subspace that improves the received SNR [10], [11]. Nevertheless, the interference due to other users can not be fully cancelled by analog beamformers because of the limitation which is fixed magnitude of its elements [3], [10]. Therefore, a low dimension digital prcoding is performed in baseband alongwith the analog prcoding in radio frequency domain and this combined analog and digital precoding is called as hybrid precoding [3].

In the conventional multi-user MISO scenario, interference between users is cancelled by zero forcing method which is inverting the matrix formed by collecting the channel matrix of interfering users [12].

Most of the current mmWave communication schemes consider only single type of MIMO wireless networks [3], [2]. In [13], authors present the idea of sharing the mmWave spectrum band. In this paper, feasibility and coordination is evaluated between two next generation mmWave networks without considering details of the analog and digital precoding.

Towards this end, based on the applications discussed in [13], several services can be allocated to a mmWave band of frequency spectrum. Therefore, it is highly imperative to restrict the interference on the necessary services of the licensed user. In this work, we consider this type of interference while designing the precoder and combiners.

Further, in the context of conventional microwave spectrum sharing systems, throughput maximization scheme is analysed in [14], [15], [16], [6]. In the seminal work of [14], the author considers only idea of throughput maximization without providing any theory of beamforming. The work in [15] demonstrates that multi-carrier technique is suitable to share the spectrum. However, only restricted number of users can coexist with primary user because of deployment of single antenna at the base station and secondary user. In [6] and [16], authors consider multiple antenna based cognitive systems and obtain the expression for throughput and mean square error (MSE). In [6], a throughput maximization problem for CR system with conventional and large system setting is presented, in which zero forcing scheme is employed. Further, relevant analysis of the asymptotic beamforming is presented in this work, however, which lacks to provide practical implementa-tion details of the system. However, these existing works [14], [15], [16], [6] do not consider the analog processing part of the precoder and combiner designing, which is a challenging problem.

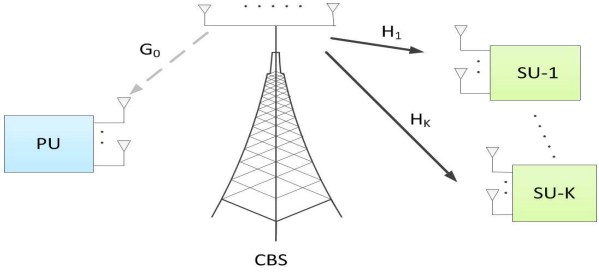

Fig. 1: System model of mmWave MIMO underlay cognitive radio systems

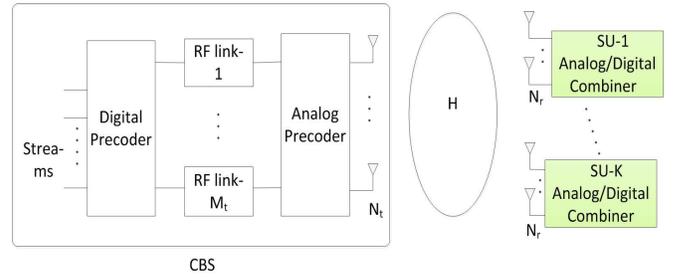

Fig. 2: Analog and digital architecture of the CBS serving multiple secondary users

### A. Contributions of this work

The contributions of this work are summarized as follows. In downlink communication of mmWave band sharing CR system, we model the mmWave MIMO channel between the CBS and secondary users as well as the interference channel between the CBS and primary user by describing antenna array steering vectors which are relied upon the angle of arrival and angle of departure parameters. For each antenna element, employing RF chain is quit expensive hardware demand. Therefore, the precoding procedure is categorized into two stages which are analog and digital stages. Analog precoding is analysed by employing a number of RF links which are less than the total number of antenna elements. The analog precoding is explained when multiple data streams are transmitted from the CBS. Further, a procedure of gathering all the unlicensed mmWave channel matrices between the CBS and secondary users and constructing analog precoder at the CBS is presented. Subsequently, analog combiner is designed by capturing the concept of received SNR maximization by exploiting large MIMO channel gains. Moreover, a scheme is developed to obtain the combining matrix of all secondary users. The computation in this scheme is possible because of smaller number of antennas at secondary user than that of the CBS. Further, after processing the analog framework, baseband processing is analysed by digital precoder which employ block diagonalization scheme to cancel the interference due to other secondary users. After that digital combining matrix is calculated which is simple to implement without any overhead of feeding the mmWave channel information back to the CBS. As the licensed primary user of mmWave band has higher priority than unlicensed secondary users, the proposed scheme governs that interference to primary users does not exceed certain threshold by allocating the transmit power based on the condition of mmWave channel to improve the throughput of the proposed system.

### B. Organization

The organization of the work is given as follows. The system model for the mmWave MIMO cognitive radio system is presented in Section II. Section III presents a detail analysis of geometrical modelling of channels in this setup. This section is proceeded by calculation of analog precoder/combiner and capturing the framework of inter-secondary user interference in the Section IV. This is followed by rate maximization scheme based on block diagonalization in the proposed scenario in Section V. Simulation results are presented to compare the performance of the proposed scheme in section VI and followed by the conclusion in section VII.

### C. Notation

The notations used throughout in this work are given as follows. All the matrices and vectors are given in upper-case and lower-case boldface letters and magnitude of a scalar is given by $| \; . \; |$. The conjugate transpose, transpose and the $(i, j)$th element of a matrix are given by $(.)^H$, $(.)^T$ and $[ \; . \; ]_{i,j}$. The notation $\mathbb{C}^{R \times T}$ and $\mathcal{D}(\mathbf{a})$ denote a matrix of size $R \times T$ with complex entries and a diagonal matrix with vector $\mathbf{a}$ on the main diagonal.

## II. SYSTEM MODEL

In this work, we consider a millimeter wave (mmWave) MIMO cognitive radio system which is operating in underlay mode. In this mode, the licensed primary system provides an opportunity to share the mmWave radio spectrum with the unlicensed secondary system, irrespective of the transmission status of primary system. However, the secondary system is not allowed to lessen the quality of received signal at the primary user below a predefined threshold level [6], [7]. Fig. 1 shows the system model of this downlink CR system where a cognitive base station (CBS) communicates with total $K$ secondary users (SU) in presence of an existing primary user (PU).

The cognitive base station is equipped with $N_t$ transmit antennas and $M_t$ RF links. This CBS sends the data signal to secondary users. Each secondary user consists $N_r$ receive antennas and $M_r$ RF links and supports $D$ streams of data as illustrated in Fig. 2. Thus, the cognitive base station communicates with total $KD$ streams of data. As shown in Fig. 2, the CBS comprises of $M_t \times KD$ dimension digital precoding matrix $\mathbf{B}$.

Let $\mathbf{H}_k \in \mathbb{C}^{N_r \times N_t}, k = 1, \ldots, K$ represents the mmWave MIMO channel matrix from the CBS to the $k$th secondary user and $\mathbf{G}_0 \in \mathbb{C}^{N_{r0} \times N_t}$ denotes the mmWave MIMO channel matrix from the CBS to the primary user. We assume that the data streams $KD$ communicated by the CBS follows the condition, $KD \leq M_t \leq N_t$. Similarly, the received streams $D$ at each secondary user follows the inequality, $D \leq M_r \leq N_r$. We employ a simple underlay CR model as in [6], [7] in which the received signal $\mathbf{y}_k \in \mathbb{C}^{N_r \times 1}$ at the $k$th secondary user is given as

$$\mathbf{y}_k = \mathbf{H}_k \mathbf{x}_k + \sum_{j=1, j \neq k}^{K} \mathbf{H}_k \mathbf{x}_j + \boldsymbol{\eta}_k, \ k = 1, \ldots, K, \quad (1)$$

where $\boldsymbol{\eta}_k \in \mathbb{C}^{N_r \times 1}$ denotes independent and identically distributed (i.i.d.) additive Gaussian noise with zero mean and unity variance. The vector $\mathbf{x}_k \in \mathbb{C}^{N_t \times 1}$ is given as

$$\mathbf{x}_k = \mathbf{FB}_k \tilde{\mathbf{x}}_k, \quad (2)$$

where $\mathbf{B}_k$ is the $M_t \times D$ dimensional matrix and $\tilde{\mathbf{x}}_k$ is the $D \times 1$ dimensional symbol vector. The matrix $\mathbf{B}_k$ is obtained as $\mathbf{B}_k = \mathbf{B} (:, ((k-1)D+1) : kD)$. Further, analog RF precoder $\mathbf{F}$ of dimension $N_t \times M_t$ is linked to the digital precoder as shown in Fig 2.

Each of the secondary user is equipped with an analog combiner $\mathbf{W}_k$, $1 \leq k \leq K$ in RF domain and followed by a digital combiner $\mathbf{T}_k$, $1 \leq k \leq K$ with dimension $N_r \times M_r$ and $M_r \times D$ respectively. The received signal $\mathbf{y}_k \in \mathbb{C}^{D \times 1}$ at the $k$th secondary user after analog and digital processing is written as

$$\tilde{\mathbf{y}}_k = \mathbf{T}_k^H \mathbf{W}_k^H \mathbf{H}_k \mathbf{x}_k + \sum_{j=1, j \neq k}^{K} \mathbf{T}_k^H \mathbf{W}_k^H \mathbf{H}_k \mathbf{x}_j + \mathbf{T}_k^H \mathbf{W}_k^H \boldsymbol{\eta}_k. \quad (3)$$

By aid of (2), the above equation can be rewritten as

$$\begin{aligned} \tilde{\mathbf{y}}_k = & \mathbf{T}_k^H \mathbf{W}_k^H \mathbf{H}_k \mathbf{FB}_k \tilde{\mathbf{x}}_k + \sum_{j=1, j \neq k}^{K} \mathbf{T}_k^H \mathbf{W}_k^H \mathbf{H}_k \mathbf{FB}_j \tilde{\mathbf{x}}_j \\ & + \mathbf{T}_k^H \mathbf{W}_k^H \boldsymbol{\eta}_k. \end{aligned} \quad (4)$$

Next, we will present the mmWave channels as a geometrical channel model similar to the work in [8].

## III. Modeling of mmWave channel between the CBS and secondary user as a geometrical MIMO channel

MmWave channels are highly directional and blockage-sensitive in nature. This leads to limited number of scatterers in the spatial path of signal while propagating from the CBS to secondary user [2], [8]. This observation of mmWave channel is modelled in (5) which is similar to the expression given in [8]. The mmWave channel associated with the $k$th secondary user is calculated by sum of all the propagation paths contributed by $L_k$ scatterers. We assume that each of the scatterer incorporates only a single spatial path while propagating from the CBS to secondary user. Therefore, the total number of spatial paths of the $k$th secondary user is also $L_k$. Thus, the MIMO channel matrix $\mathbf{H}_k$ can be formulated as

$$\mathbf{H}_k = \sqrt{\frac{N_t N_r}{L_k}} \sum_{l=1}^{L_k} \alpha_{k,l} \mathbf{a}_r (\theta_{k,l}) \mathbf{a}_t^H (\phi_{k,l}), \quad (5)$$

where $\alpha_{k,l} \sim \mathcal{CN}(0, \sigma_\alpha^2)$ denotes the complex valued gain of the $l$th path associated with the $k$th secondary user. The quantity $\theta_{k,l}$ is angle of arriving (AoA) of plane wave propagating through the $l$th path associated with the $k$th secondary user. Similarly, quantity $\phi_{k,l}$ is angle of departing (AoD) of plane wave which is propagating through the $l$th path for the $k$th secondary user. The quantities $\theta_{k,l}$ and $\phi_{k,l}$ lie in the interval $[0, 2\pi]$. Further, vectors $\mathbf{a}_r (\theta_{k,l})$ and $\mathbf{a}_t (\phi_{k,l})$ are antenna array steering vectors of the $k$th secondary user and the CBS respectively. We consider uniform linear antenna array (ULA) model. Therefore, the array steering vectors $\mathbf{a}_t (\phi_{k,l})$ at the CBS and $\mathbf{a}_r (\theta_{k,l})$ at the $k$th secondary user are given as

$$\mathbf{a}_t (\phi_{k,l}) = \frac{1}{\sqrt{N_t}} \left[ 1, e^{j \frac{2\pi}{\lambda} d_t \sin(\phi_{k,l})}, \ldots, e^{j(N_t-1) \frac{2\pi}{\lambda} d_t \sin(\phi_{k,l})} \right]^T,$$
$$\mathbf{a}_r (\theta_{k,l}) = \frac{1}{\sqrt{N_r}} \left[ 1, e^{j \frac{2\pi}{\lambda} d_r \sin(\theta_{k,l})}, \ldots, e^{j(N_r-1) \frac{2\pi}{\lambda} d_r \sin(\theta_{k,l})} \right]^T.$$
$$(6)$$

where $d_t, d_r$ are the separation distances between nearest antenna elements at the CBS and at the secondary user. Further, $\lambda$ is the wavelength of the mmWave signal.

In the mmWave based wireless systems, the number of RF links are kept lower than the number of antennas [3]. When the number of RF links are as large as number of antennas, this yields high cost while deploying conventional precoding schemes discussed in [12]. One promising approach of reducing the cost is keeping the number of RF links below the number of antennas. Hence, we employ the condition $M_t < N_t$ and $M_r < N_r$. However, one can exploit the significantly large antenna array gain by the aid of the analog combiner and analog precoder.

## IV. Analog and digital precoder at the CBS and combiner at secondary users (ADPC)

In this section, we focus on developing strategies for analog precoding and combining for the above framework. The channel matrix $\tilde{\mathbf{H}}_k \in \mathbb{C}^{M_r \times M_t}$ associated with the $k$th secondary user is defined as

$$\tilde{\mathbf{H}}_k = \mathbf{W}_k^H \mathbf{H}_k \mathbf{F}, \ k = 1, \ldots, K. \quad (7)$$

From the above (7), the received signal expression (4) can be further simplified as

$$\tilde{\mathbf{y}}_k = \mathbf{T}_k^H \tilde{\mathbf{H}}_k \mathbf{B}_k \mathbf{x}_k + \sum_{j=1, j \neq k}^{K} \mathbf{T}_k^H \tilde{\mathbf{H}}_k \mathbf{B}_j \mathbf{x}_j + \mathbf{T}_k^H \mathbf{W}_k^H \boldsymbol{\eta}_k. \quad (8)$$

The equivalent channel matrix $\hat{\mathbf{H}} \in \mathbb{C}^{KM_r \times M_t}$ in base-band domain is obtained by concatenating the matrix $\tilde{\mathbf{H}}_k, 1 \leq k \leq$

$K$ of all the secondary users which can be expressed as

$$\hat{\mathbf{H}} = \begin{bmatrix} \hat{\mathbf{H}}_1 \\ \vdots \\ \hat{\mathbf{H}}_K \end{bmatrix} = \begin{bmatrix} \mathbf{W}_1^H & \cdots & \mathbf{0} \\ \vdots & \ddots & \vdots \\ \mathbf{0} & \cdots & \mathbf{W}_K^H \end{bmatrix} \begin{bmatrix} \mathbf{H}_1 \\ \vdots \\ \mathbf{H}_K \end{bmatrix} \mathbf{F}. \quad (9)$$

From the above analysis, one can observe that $\hat{\mathbf{H}}$ denotes the matrix corresponding to the base-band equivalent channel formed between the CBS and all the secondary users. Next, we develop the design of analog precoder at the CBS assuming that analog combining matrices are already given. Nevertheless, the detail computation procedure of these combining matrices are described in later part of this section.

The analog precoder $\mathbf{F}$ at the CBS is designed to control the phase angle $\varphi_{i,j}$ at the radio frequency domain. Further, the matrix $\check{\mathbf{H}}$ with dimension $KM_r \times N_t$ is defined as

$$\check{\mathbf{H}} = \begin{bmatrix} \mathbf{W}_1^H \mathbf{H}_1 \\ \vdots \\ \mathbf{W}_K^H \mathbf{H}_K \end{bmatrix} = \begin{bmatrix} \check{h}_{1,1} & \cdots & \check{h}_{1,N_t} \\ \vdots & & \vdots \\ \check{h}_{KM_r,1} & \cdots & \check{h}_{KM_r,N_t} \end{bmatrix}. \quad (10)$$

The phase angles $\varphi_{i,j}, \ 1 \le i \le N_t, 1 \le j \le KM_r$ are calculated from the phase of each entry of a matrix $\check{\mathbf{H}}^H$. Thus, the angle $\varphi_{i,j}$ is given as $\varphi_{i,j} = \check{h}_{i,j}$. Finally, the $(i,j)$th entry of the analog precoding matrix $\mathbf{F}$ is computed as

$$[\mathbf{F}]_{i,j} = \frac{e^{-j\varphi_{i,j}}}{\sqrt{N_t}} \ 1 \le i \le N_t, 1 \le j \le KM_r. \quad (11)$$

The computation of $\mathbf{F}$ requires total $KM_r$ number of radio frequency links at the CBS which is similar to work in [3]. Therefore, the dimension of $\mathbf{F}$ becomes $N_t \times KM_r$ which means $M_t = KM_r$. Thus, the matrix $\check{\mathbf{H}}$ comprises of equal number of rows and columns, and hence, it has to be a square matrix with the $KM_r \times KM_r$ dimension. In the following section, we assume that the matrix $\check{\mathbf{H}}$ is not rank-deficient in nature. This supports transmission of data through multiple number of streams without distorted by noise amplification at the secondary users [12].

To cater the benefits of high throughput, each transmitted data stream requires large antenna diversity which is exploited by antenna arrays at the CBS and secondary users [11], [3]. Next, we design analog combiner at the secondary user which is capable of adjusting the phase with moderate level of hardware requirements. The analog combining matrix $\mathbf{W}_k$ at the $k$th secondary user is computed by processing the square of the element of the main diagonal of matrix $\hat{\mathbf{H}}$.

Let $\mathbf{w}_{k,m}$ denotes the $m$th column of $\mathbf{W}_k$ which is associated with the $m$th RF link of the $k$th secondary user. As discussed above, equivalent matrix $\hat{\mathbf{H}}$ is a square matrix and hence, the $i$th index of its main diagonal can be written as

$$i = (k-1)M_r + m, 1 \le k \le K, 1 \le m \le M_r. \quad (12)$$

The above expression of the $i$th index can be utilize in the following relation associated with $\hat{\mathbf{H}}$ as given below

$$\left[\hat{\mathbf{H}}\right]_{i,i} = \left[\hat{\mathbf{H}}\right]_{(k-1)M_r+m,(k-1)M_r+m}, \quad (13)$$

where $\left[\hat{\mathbf{H}}\right]_{i,i}$ denotes the $i$th element of the main diagonal of $\hat{\mathbf{H}}$. By aid of (10) and (9), one can compute the element $\left[\hat{\mathbf{H}}\right]_{i,i}$ as

$$\left[\hat{\mathbf{H}}\right]_{i,i} = \left[\check{h}_{i,1}, \check{h}_{i,2}, \ldots, \check{h}_{i,N_t}\right]$$
$$\times \begin{bmatrix} [\mathbf{F}]_{1,i} \\ \vdots \\ [\mathbf{F}]_{N_t,i} \end{bmatrix}$$
$$= \sum_{j=1}^{N_t} \check{h}_{i,j} [\mathbf{F}]_{j,i}$$
$$= \frac{1}{\sqrt{N_t}} \sum_{j=1}^{N_t} \check{h}_{i,j} e^{-j\varphi_{j,i}}$$
$$= \sum_{j=1}^{N_t} |\check{h}_{i,j}|, \quad (14)$$

where $|\check{h}_{i,j}|$ is the magnitude of the $j$th element of the $i$th row vector $\check{\mathbf{h}}_i$ of the matrix $\check{\mathbf{H}}$.

Now, we need to find the expression of vector $\check{\mathbf{h}}_i$ in terms of $\mathbf{w}_{k,m}^H$ and $\mathbf{h}_{k,j}$. For the $k$th secondary user, the dimensions of the $m$th row vector $\mathbf{w}_{k,m}^H$ of $\mathbf{W}_k^H$ and the $j$th column vector $\mathbf{h}_{k,j}$ of $\mathbf{H}_k$ are $1 \times N_r$ and $N_r \times 1$ respectively. By aid of (10) and (12), the $i$th row vector $\check{\mathbf{h}}_i \in \mathbb{C}^{1 \times N_t}$ can be observed as

$$\check{\mathbf{h}}_i = \left[\mathbf{w}_{k,m}^H \mathbf{h}_{k,1} \ldots \mathbf{w}_{k,m}^H \mathbf{h}_{k,j} \ldots \mathbf{w}_{k,m}^H \mathbf{h}_{k,N_t}\right] \quad (15)$$

Therefore, one can further simplify (14) as

$$\left[\hat{\mathbf{H}}\right]_{i,i} = \sum_{j=1}^{N_t} |\check{h}_{i,j}| = \sum_{j=1}^{N_t} |\mathbf{w}_{k,m}^H \mathbf{h}_{k,j}|. \quad (16)$$

Our next concern is maximizing the quantity $\sum_{k=1}^{K} \sum_{m=1}^{M_r} \sum_{j=1}^{N_t} |\mathbf{w}_{k,m}^H \mathbf{h}_{k,j}|^2$ to maximize the receive signal to noise ratio (SNR) [17], [3]. In the above setting, we consider independent design of $\mathbf{W}_k$ for each of the secondary user. Therefore, the above maximization is equivalent to maximization of only $\sum_{m=1}^{M_r} \sum_{j=1}^{N_t} |\mathbf{w}_{k,m}^H \mathbf{h}_{k,j}|^2$. Hence, analog combining matrix $\mathbf{W}_k$ at the $k$th secondary user can be computed by the following problem

$$\max \sum_{m=1}^{M_r} \sum_{j=1}^{N_t} |\mathbf{w}_{k,m}^H \mathbf{h}_{k,j}|^2, \ 1 \le k \le K$$

$$\text{s.t. } |[\mathbf{W}_k]_{i,j}| = \left(\sqrt{N_r}\right)^{-1}, \ 1 \le i \le N_r, 1 \le j \le M_r. \quad (17)$$

The equality condition in above (17) reflects the constant magnitude restriction on each of the element of the matrix $\mathbf{W}_k$ because of only phase angle can be controlled in the analog combiner. The above problem is intractable and does not provide any optimal solution in the closed form [3], [18]. The problem (17) is difficult to solve because of the non-convex feasibility constraint which is the constant magnitude elements of the matrices $\mathbf{W}_k$. To provide a simple efficient solution to the above problem (17), one can employ the characteristics of geometrical model of mmWave channel which is discussed in (5). Interestingly, this geometrical model can be exploited along with the projection of $\mathbf{w}_{k,m}$ on $\mathbf{h}_{k,j}$ to find the solution. The procedure is given as follows.

We know that the columns of optimal unitary combiner matrix forms an orthonormal basis for the channel matrix's column space [17]. Further, the column space of the channel matrix $\mathbf{H}_k$ is spanned by the vectors $\mathbf{a}_r\left(\theta_{k,l}\right), 1 \leq l \leq L_k$ as defined in (5). Thus, the column vectors $\mathbf{h}_{k,j}, 1 \leq j \leq N_t$ of $\mathbf{H}_k$ are the combination of the vectors $\mathbf{a}_r\left(\theta_{k,l}\right)$ which forms the basis of column space of $\mathbf{H}_k$ when the total number of spatial paths $L_k$ follows the condition $L_k \leq N_r$ [3]. Next, consider that the angles $\theta_{k,l}, 1 \leq k \leq K, 1 \leq l \leq L_K$ are recognizably different angles of arrivals for all the secondary users. Therefore, all the vectors $\mathbf{a}_r\left(\theta_{k,l}\right)$ associated with the $k$th secondary user are independent antenna array response vectors. Hence, the channel matrix $\mathbf{H}_k$ formed by the aid of these $\mathbf{a}_r\left(\theta_{k,l}\right)$ vectors is a full rank matrix.

Proceeding with the above analysis, $\mathbf{w}_{k,m}$ the columns of $\mathbf{W}_k$ are obtained from the linear combinations of $\mathbf{a}_r\left(\theta_{k,l}\right), 1 \leq l \leq L_k$. Therefore, $|\mathbf{w}_{k,m}^H \mathbf{h}_{k,j}|$ is maximized by projecting $\mathbf{a}_r\left(\theta_{k,l}\right)$ on $\mathbf{h}_{k,j}$. Towards this end, we define a vector $\mathbf{a}\left(\zeta_k\right)$ as

$$\mathbf{a}\left(\zeta_k\right) = \left[1, e^{j\zeta_k}, e^{j2\zeta_k}, \dots, e^{j(N_r-1)\zeta_k}\right]^T / \sqrt{N_r}, \quad (18)$$

where the angle $\zeta_k$ is defined as $\zeta_k = \frac{2\pi d}{\lambda}\sin\theta_k$. Next, the angle $\zeta_k$ is discretized into an $N_r$ number of states over the interval $[0, 2\pi]$ and therefore, $\zeta^i$ which is the $i$th discretized value of $\zeta_k$ is given as

$$\zeta^i = \frac{2\pi(i-1)}{N_r}, i = 1, 2, \dots, N_r \quad (19)$$

The quantities $\zeta^i$ are utilized to form a set $\mathbf{A}$ of the vectors $\mathbf{a}\left(\zeta^i\right)$ which is expressed as

$$\mathbf{A} = \left\{\mathbf{a}\left(0\right), \mathbf{a}\left(\frac{2\pi}{N_r}\right), \dots, \mathbf{a}\left(\frac{2\pi(N_r-1)}{N_r}\right)\right\}. \quad (20)$$

Since the vectors $\mathbf{a}\left(\zeta^i\right)$ can be efficiently leveraged to construct the columns of $\mathbf{W}_k$, the analog combiner designing problem (14) can be readily expressed as

$$\max_{\mathbf{w}_{k,m}} \sum_{m=1}^{M_r} \sum_{j=1}^{N_t} |\mathbf{w}_{k,m}^H \mathbf{h}_{k,j}|^2, \ 1 \leq k \leq K$$
$$\text{s.t. } \mathbf{w}_{k,m} \in \left\{\mathbf{a}\left(\zeta^i\right), 1 \leq i \leq N_r\right\}. \quad (21)$$

Now, one can note that the analog combiner designing problem becomes tractable because of re-modelling the constraint which is selecting the vector $\mathbf{w}_{k,m}$ from the set of $N_r$ number of $\mathbf{a}\left(\zeta^i\right)$ vectors.

The procedure to solve the above problem is given in detail as follows. Firstly, $|\mathbf{a}\left(\zeta^i\right)^H \mathbf{h}_{k,j}|$ is computed, and then added for $j = 1, \dots, N_t$. Secondly, all these resulting $N_r$ terms corresponding to the set $\left\{\mathbf{a}\left(\zeta^i\right), 1 \leq i \leq N_r\right\}$ are arranged from the largest to the smallest in magnitude. Consequently, we select the first $M_r$ number of vectors $\mathbf{a}\left(\zeta^i\right)$. Finally, these selected vectors are employed as the columns of the matrix $\mathbf{W}_k$ to maximize the objective function of (21). The above procedure is simple brute force technique similar to [11].

In order to null the interference between secondary users,

one can derive the expression of digital precoder and combiner based on block diagonalization methodology [12] for mmWave based CR system. Following the above scheme, analog precoding matrix and combining matrices are obtained as $\mathbf{F}$ and $\mathbf{W}$ respectively. For the $k$th secondary user, equivalent matrix $\bar{\mathbf{H}}_k$ in base-band domain is calculated in (9) and then it is processed further by null space projection technique. We begin by considering the $(K-1)M_r \times M_t$ matrix $\tilde{\mathbf{H}}_k$ which is obtained by stacking $(K-1)$ rest of the secondary users' channel matrices as given below

$$\tilde{\mathbf{H}}_k = \left[\bar{\mathbf{H}}_1^T, \bar{\mathbf{H}}_2^T, \dots, \bar{\mathbf{H}}_{k-1}^T, \bar{\mathbf{H}}_{k+1}^T, \dots, \bar{\mathbf{H}}_K^T\right]^T. \quad (22)$$

Let $\bar{F}_k$ denotes the rank of $\tilde{\mathbf{H}}_k$ and follows the condition $\bar{F}_k < M_t$. Further, the nullity of $\tilde{\mathbf{H}}_k$ is considered as $M_r$. The singular value decomposition (SVD) is employed to yield the digital precoding matrix $\mathbf{B}_k$ as follows. The SVD of $\tilde{\mathbf{H}}_k$ can be readily written as

$$\tilde{\mathbf{H}}_k = \bar{\mathbf{U}}_k \bar{\mathbf{\Sigma}}_k \left[\bar{\mathbf{V}}_k^{(1)} \ \bar{\mathbf{V}}_k^{(0)}\right]^H, \quad (23)$$

where $\bar{\mathbf{V}}_k^{(1)}$ and $\bar{\mathbf{V}}_k^{(0)}$ are of dimensions $M_t \times (M_t - M_r)$ and $M_t \times M_r$ respectively. The matrix $\bar{\mathbf{V}}_k^0$ comprises of the last $M_r$ right singular vectors of $\tilde{\mathbf{H}}_k$. Thus, the columns of matrix $\bar{\mathbf{V}}_k^0$ span the null space of $\tilde{\mathbf{H}}_k$ and therefore, exploited in the computation of $\mathbf{B}_k$. The matrix $\bar{\mathbf{H}}_k \bar{\mathbf{V}}_k$ associated with the $k$th secondary user signifies an effective channel without getting interfered from the other $K-1$ secondary users. Let us define the block diagonal channel matrix $\acute{\mathbf{H}} \in \mathbb{C}^{KM_r \times KM_r}$ in base-band domain as

$$\acute{\mathbf{H}} = \begin{bmatrix} \bar{\mathbf{H}}_1\bar{\mathbf{V}}_1^0 & \cdots & \mathbf{0} \\ \vdots & \ddots & \vdots \\ \mathbf{0} & \cdots & \bar{\mathbf{H}}_K\bar{\mathbf{V}}_K^0 \end{bmatrix}. \quad (24)$$

It can be readily observed that the above block diagonal (BD) channel matrix $\acute{\mathbf{H}}$ corresponds to equivalent channel matrix which nulls the inter-secondary user interference. As discussed previously, $\acute{\mathbf{H}}$ is not a rank deficient in nature and $\mathbf{H}_k$ is a full rank channel matrix. Hence, it follows that the columns of the $M_r \times M_r$ dimensional matrix $\bar{\mathbf{H}}_k\bar{\mathbf{V}}_k^0$ are linearly independent. Therefore, one can infer that the CBS facilitates the transmission of $M_r \geq D$ number of streams. Hence, the rank of $\bar{\mathbf{H}}_k\bar{\mathbf{V}}_k^0$ is $M_r$. Now, consider the SVD of the $M_r \times M_r$ matrix $\bar{\mathbf{H}}_k\bar{\mathbf{V}}_k^0$ which is given as

$$\bar{\mathbf{H}}_k\bar{\mathbf{V}}_k^0 = \mathbf{U}_k\mathbf{\Sigma}_k\mathbf{V}_k^H, \quad (25)$$

where $\mathbf{\Sigma}_k$ denotes a diagonal matrix with the singular values $\sigma_{k,i} \ 1 \leq i \leq M_r$ in the main diagonal, and the dimensions of matrices $\mathbf{\Sigma}_k, \mathbf{U}_k$ and $\mathbf{V}_k$ are $M_r \times M_r$. Further, the matrix $\mathbf{V}_k$ can be partitioned into sub-matrices as given below

$$\mathbf{V}_k = \left[\mathbf{V}_k^{(1)} \ \mathbf{V}_k^{(2)}\right], \quad (26)$$

where $\mathbf{V}_k^{(1)} = \mathbf{V}_k\left(:, 1:D\right)$ is the $M_r \times D$ dimensional matrix. Similarly, one can write the expression of $\mathbf{U}_k$ as

$$\mathbf{U}_k = \left[\mathbf{U}_k^{(1)} \ \mathbf{U}_k^{(2)}\right], \quad (27)$$

where $\mathbf{U}_k^{(1)} = \mathbf{U}_k(:, 1 : D)$ is the $M_r \times D$ dimensional matrix. Towards this end, one can compute the digital precoding matrix $\mathbf{B}_k \in \mathbb{C}^{M_t \times D}$ corresponding to the $k$th secondary user as

$$\mathbf{B}_k = \bar{\mathbf{V}}_k^{(0)} \mathbf{V}_k^{(1)}. \tag{28}$$

As a result of (25) and (27), the digital combining matrix $\mathbf{T}_k$ corresponding to the $k$th secondary user can be given as

$$\mathbf{T}_k = \mathbf{U}_k^{(1)}. \tag{29}$$

In the proposed cognitive scenario, the achievable sum-rate by the mmWave MIMO channel based scheme is given as

$$C = \sum_{k=1}^{K} \sum_{d=1}^{D} \log_2 \left(1 + P_{k,d} \sigma_{k,d}^2\right), \tag{30}$$

where $P_{k,d}$ is the $i$th diagonal element of $D \times D$ diagonal power allocation matrix $\mathcal{D}(\mathbf{p}_k)$. Note that in the above (30), we use the fact that analog combining matrix $\mathbf{W}_k$ comprises of the columns which are orthogonal to each other and therefore $\mathbf{W}_k^H \mathbf{W}_k = \mathbf{I}_{M_r}$. Further, one can see that $\mathbf{T}_k^H \mathbf{T}_k = \mathbf{I}_D$.

## V. ANALOG AND DIGITAL PRECODING/COMBINING BASED RATE MAXIMIZATION

After analog and digital processing to design precoder and combiners, the sum-rate maximization problem for mmWave MIMO channel based cognitive radio system is addressed in this section. The power received at the primary user due to transmission by the CBS is kept below a threshold value. The covariance matrix corresponding to interference signal received at the primary user because of communication between the CBS and all the secondary users is formulated as

$$\mathbf{J}_0 = \sum_{k=1}^{K} \mathbf{G}_0 \mathbf{R}_k \mathbf{G}_0^H, \tag{31}$$

where $\mathbf{R}_k \in \mathbb{C}^{N_t \times N_t}$ is given as $\mathbf{R}_k = \mathbf{FB}_k \mathcal{D}(\mathbf{p}_k) \mathbf{B}_k^H \mathbf{F}^H$. Furthermore, $\mathbf{G}_0$ can also be considered as geometrically modeled channel similar to analysis given in Section (III). To limit this interference power received at the primary user, one can formulate the spatial interference constraint which is expressed as

$$\sum_{k=1}^{K} \text{Tr}\left(\mathbf{G}_0 \mathbf{R}_k \mathbf{G}_0^H\right) \leq I_{th}, \tag{32}$$

where $I_{th}$ denotes threshold value of interference received at primary user due to communication in secondary system. Further, the received interference power at primary user can not be greater than this $I_{th}$ value. The quantity $\text{Tr}\left(\mathbf{G}_0 \mathbf{R}_k \mathbf{G}_0^H\right)$

can be further simplified as

$$\text{Tr}\left(\mathbf{G}_0 \mathbf{R}_k \mathbf{G}_0^H\right) = \text{Tr}\left(\mathbf{B}_k \mathcal{D}(\mathbf{p}_k) \mathbf{B}_k^H \mathbf{F}^H \mathbf{G}_0^H \mathbf{G}_0 \mathbf{F}\right)$$

$$= \text{Tr}\left(\mathcal{D}(\mathbf{p}_k) \underbrace{\mathbf{B}_k^H \mathbf{F}^H \mathbf{G}_0^H \mathbf{G}_0 \mathbf{F} \mathbf{B}_k}_{\boldsymbol{\Gamma}_k}\right)$$

$$= \sum_{d=1}^{D} P_{k,d} \gamma_{k,d}, \tag{33}$$

where $P_{k,d}, \gamma_{k,d}$ are the $d$th element of principal diagonal of $\mathcal{D}(\mathbf{p}_k)$ and $\boldsymbol{\Gamma}_k$ respectively. Therefore, spatial interference constraint (32) can be rewritten as

$$\sum_{k=1}^{K} \sum_{d=1}^{D} P_{k,i} \gamma_{k,i} \leq I_{th}. \tag{34}$$

Similar to the work in [15], we consider that the CBS has enough power supply. The sum-rate maximization (SRM) problem for the mmWave MIMO channel based CR system is given as

$$\max_{P_{k,d}} \sum_{k=1}^{K} \sum_{d=1}^{D} \log_2 \left(1 + P_{k,d} \sigma_{k,d}^2\right)$$

$$\text{s.t. } \sum_{k=1}^{K} \sum_{d=1}^{D} P_{k,d} \gamma_{k,d} \leq I_{th}$$

$$P_{k,d} \geq 0, \ 1 \leq k \leq K, 1 \leq d \leq D. \tag{35}$$

Employing the above mmWave based framework, the optimal power allocation for this setting is now given by the result below.

**Theorem 1.** *The optimal power allocation for the RF phase control based analog precoding/combining and the BD based digital precoding/combining to achieve maximum sum-rate in above mmWave MIMO CR system can be given as*

$$P_{k,d} = \begin{cases} (\lambda \gamma_{k,d})^{-1} - \left(\sigma_{k,d}^2\right)^{-1}, & \text{if } \lambda < \frac{\sigma_{k,d}^2}{\gamma_{k,d}} \\ 0, & \text{otherwise} \end{cases}$$

*for $1 \leq k \leq K, 1 \leq d \leq D$.* $\tag{36}$

*Proof.* One can observe that the above problem is a non-convex in nature. Therefore, optimal solution is difficult to find. However the non-convex objective function can be written as an equivalent convex function which is given as $-\sum_{k=1}^{K} \sum_{d=1}^{D} \log_2 \left(1 + P_{k,d} \sigma_{k,d}^2\right)$. Therefore, we minimize the quantity $-\sum_{k=1}^{K} \sum_{d=1}^{D} \log_2 \left(1 + P_{k,d} \sigma_{k,d}^2\right)$ by aid of KKT framework as follows. Let $\lambda, \mu_{k,d}, 1 \leq k \leq K, 1 \leq d \leq D$ be the Lagrange coefficients associated with spatial interference constraint and power constraints in the above optimization problem (35). To solve the constrained optimization problem, we follow the standard procedure same as discussed

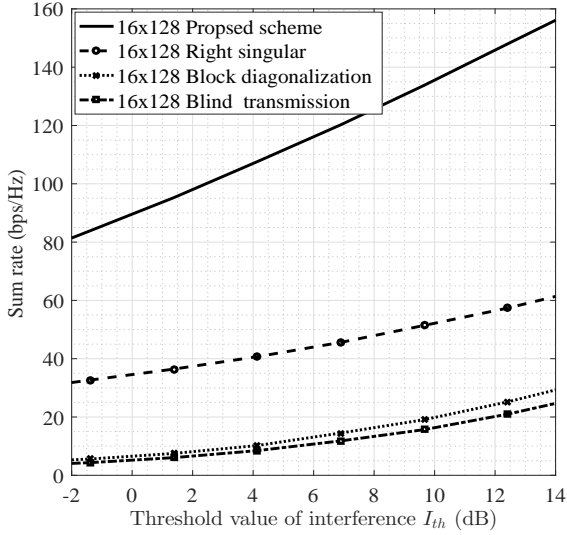

Fig. 3: Sum rate of the MIMO CR systems vs $I_{th}$ for $N_r = 16$, $N_t = 128$, $K = 8$, $M_r = N_r$, $M_t = N_t$, $N_s = 2$ in Rayleigh fading channels and comparison with various schemes

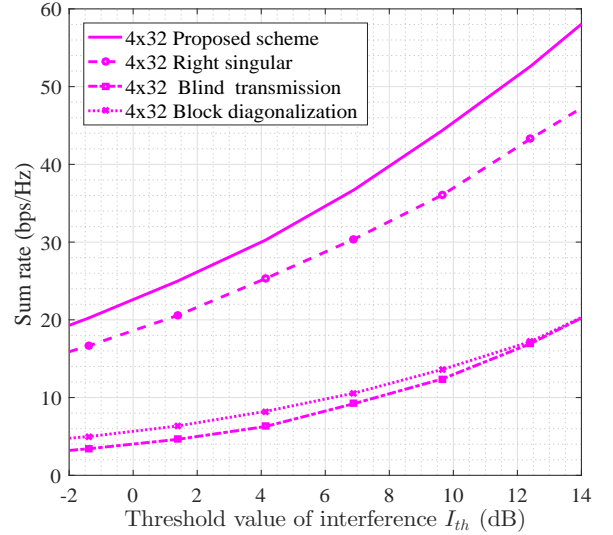

Fig. 4: Sum rate of the MIMO CR systems vs $I_{th}$ for $N_r = 4$, $N_t = 32$, $K = 8$, $M_r = N_r$, $M_t = N_t$, $N_s = 2$ in Rayleigh fading channels and comparison with various schemes

in [19].

$$p_{k,d} \geq 0 \ \mu_{k,d} \geq 0, \mu_{k,d} p_{k,d} = 0 \ \forall \, k, d$$

$$\sum_{k=1}^{K} \sum_{d=1}^{D} P_{k,d} \gamma_{k,d} \leq I_{th}$$

$$- \frac{\sigma_{k,d}^2}{1 + P_{k,d} \sigma_{k,d}^2} - \mu_{k,d} + \lambda \gamma_{k,d} = 0 \forall \, k, d \quad (37)$$

The last KKT condition is further simplified as given below.

$$- \frac{\sigma_{k,d}^2 P_{k,d}}{1 + P_{k,d} \sigma_{k,d}^2} - \mu_{k,d} P_{k,d} + \lambda \gamma_{k,d} P_{k,d} = 0$$

$$- \frac{\sigma_{k,d}^2}{1 + P_{k,d} \sigma_{k,d}^2} + \lambda \gamma_{k,d} = 0$$

$$\frac{\sigma_{k,d}^2}{\gamma_{k,d} \left( 1 + P_{k,d} \sigma_{k,d}^2 \right)} \leq \lambda \quad (38)$$

If $\lambda < \frac{\sigma_{k,d}^2}{\gamma_{k,d}}$, the above inequality holds true, if $P_{k,d} > 0$ as $\gamma_{k,d} > 0$. Thus $P_{k,d}$ can be obtained as $P_{k,d} = (\lambda \gamma_{k,d})^{-1} - \left( \frac{\sigma_{k,d}^2}{\sigma_{k,d}^2} \right)^{-1}$. However, if $\lambda \geq \frac{\sigma_{k,d}^2}{\gamma_{k,d}}$, then $P_{k,d} > 0$ is impossible because then it yields $\lambda \geq \frac{\sigma_{k,d}^2}{\gamma_{k,d} \left( 1 + P_{k,d} \sigma_{k,d}^2 \right)}$ which violates the above inequality. Hence, $P_{k,d} = 0$ if $\lambda \geq \frac{\sigma_{k,d}^2}{\gamma_{k,d}}$ holds true. $\qquad \square$

Hence, non-zero power is allocated on the $d$th stream for the $k$th secondary user when the ratio of $\sigma_{k,d}^2$ to $\gamma_{k,d}$ is greater than $\lambda$ which regulates the interference power to authorized mmWave band primary user.

## VI. SIMULATION RESULTS

To demonstrate the performance of the shared mmWave spectrum system and to analyse mmWave CR scenario discussed above. The comparison of proposed scheme with the existing schemes are presented when the propagating channels between the CBS and primary/secondary users are considered as a Rayleigh fading and geometrical model [8]. We begin with the analysis of large dimension CR systems with Rayleigh fading channels. For this scenario, the parameters are set as, $M_r = 2$ and $M_t = 16$.

Figure 3 demonstrates the sum-rate performance of large dimension CR systems in which each of the secondary user is equipped with $N_r = 16$ receive antennas and the CBS is equipped with $N_t = 128$ transmit antennas i.e. $16 \times 128$ MIMO CR system. It can be seen therein that the plot corresponding to the proposed scheme demonstrates the superior performance compared with other schemes. Our ADPC based prosed scheme performs better than the block diagonalization based scheme discussed in [20], because of analog and digital processing which exploits larger antenna array gain. Next, our scheme is compared with a recent scheme discussed in [21], in which the right singular matrix of $\mathbf{H}_k$ is employed as a precoding matrix. One can observe lower level of performance of this scheme owing to the fact that interference among secondary users is reached to a higher level. Finally, blind transmission based scheme similar to [13], in which precoding has been performed without any co-ordination with secondary users, is compared in this setting. Naturally, it results in a significant degradation in the rate performance.

For Fig. 5- Fig. 8, we consider the uniform linear array (ULA) with $d_t = d_r = \frac{\lambda}{2}$ in mmWave CR system . The number of multi-paths are limited and total number of multi-paths for all the secondary users are set as $L_k = 3$ similar to [8]. Further, variance $\sigma_\alpha^2$ of each of the path is set to unity. The angles of arrival and departure are modelled as azimuthal angles with uniform distribution in the interval $[0, 2\pi]$. In Fig. 5 and Fig. 6, the rate performances are demonstrated for $N_s = 2$ streams and $M_r = 2, M_t = 16$ RF links in $16 \times 128$, $4 \times 32$ MIMO CR systems. Fig. 5 shows that the mmWave MIMO nature of the channels due to RF links significantly improves the achievable rate performance compared to the benchmark works. Moreover, this improvement increases with increase in threshold values of $I_{th}$. Therefore, it allows increase in tolerable received interference power at primary user. These plots clearly illustrate the fact that the proposed ADPC based scheme yields a significant improvement in comparison to the scheme in which processing is based on the right singular matrix of mmWave channel similar to [21]. Next, the performance of our scheme is studied with existing schemes discussed in [20], [13]. In the context of proposed work, all the schemes are dependent on $I_{th}$ and improved by increase in number of antennas. However, the mmWave CR system achieves a non-negligible improvement than the advanced block diagonalization [20] and blind transmission schemes [13]. Similar observations can be made for $4 \times 32$ mmWave MIMO systems as shown in Fig. 6. These results also validate that the above analysis is general and still applicable to small dimensional systems.

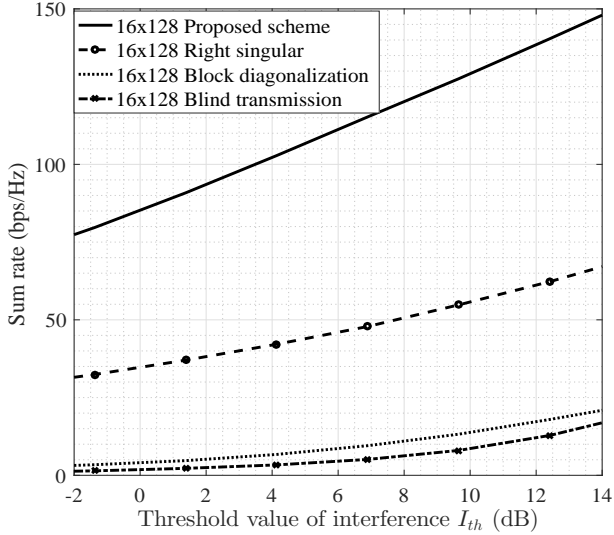

Fig. 5: Sum rate of the MIMO CR systems vs $I_{th}$ for $N_r = 16$, $N_t = 128$, $K = 8$, $M_r = 2$, $M_t = 16$, $N_s = 2$ in mmWave channels and comparison with various schemes

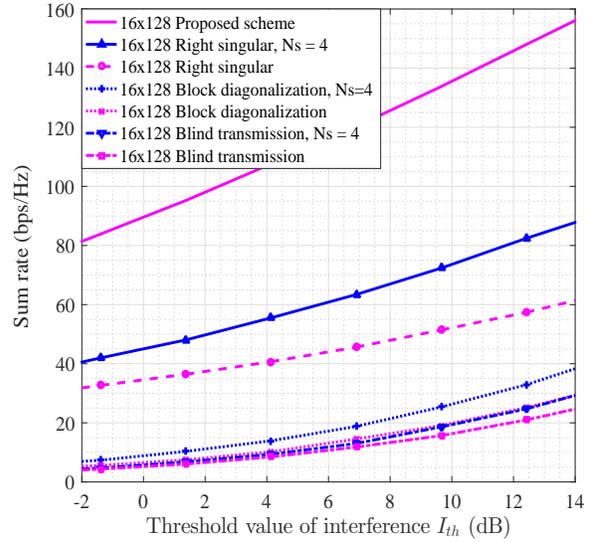

Fig. 7: Sum rate of the mmWave MIMO CR systems vs $I_{th}$ for $N_r = 16$, $N_t = 128$, $K = 8$, $M_r = N_s$, $M_t = KN_s$, $N_s = 2$ and comparison with various schemes with different $N_s$

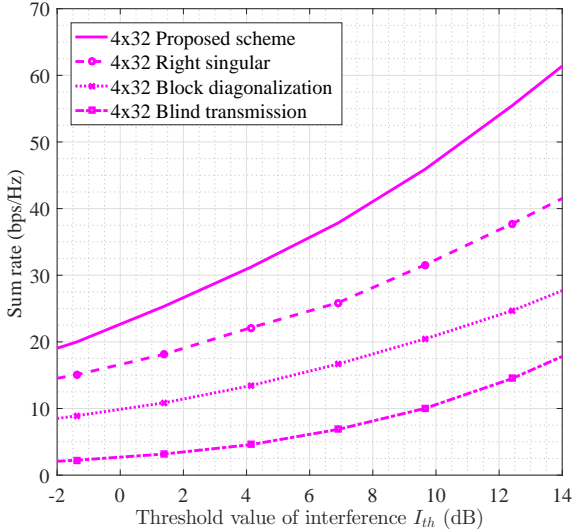

Fig. 6: Sum rate of the MIMO CR systems vs $I_{th}$ for $N_r = 4$, $N_t = 32$, $K = 8$, $M_r = 2$, $M_t = 16$, $N_s = 2$ in mmWave channels and comparison with various schemes

In Fig. 7, the rate performances are demonstrated for $N_s = 2, N_s = 4$ streams in $16 \times 128$ MIMO CR systems. Fig 7 demonstrates the relation of RF links to the sum-rate performance. The number of RF links $M_r$ employed by each secondary users are set equal to the number of streams $N_s$ to meet the cost effective multiplexing requirement i.e. $N_s \leq M_r$ of analog/digital combining. With smaller number of RF links, proposed scheme performs better than the benchmark works [21], [20], [13]. In conventional base-band CR systems, number of RF links is same as number of antenna elements, which means 16 and 128 RF links at secondary users and the CBS respectively. However, our scheme only requires 2 and 16 RF links at secondary user and the CBS respectively. Further, it can be seen that when number of streams of the benchmark works are increased the performance of our scheme is still better than others' performance due to collecting large gains of antenna arrays. Further, these array gains severely degrades the quality of interfering channel between the CBS and primary user which naturally limits the received power at primary user.

With $N_s = M_r = 4$, $I_{th} = 12$ dB settings and same settings as previous Fig. 7, next Fig. 8 demonstrates the sum-rate performance of proposed scheme when number of secondary users are increased. It can be seen that sum-rate of our scheme is higher than that of others' scheme owing to increase in diversity of multiple secondary users.

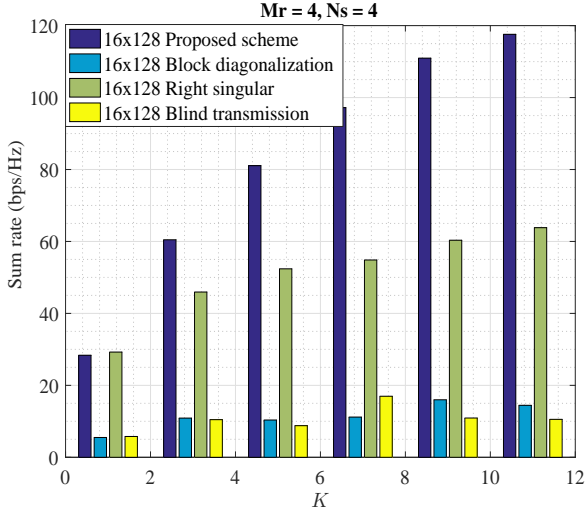

Fig. 8: Sum rate of the MIMO CR systems vs $K$ for $N_r = 16$, $N_t = 128$, $K = 8$, $M_r = N_s = 4$, $M_t = KN_s$ in mmWave channels and comparison with various schemes

## VII. Conclusion

In this work, we demonstrated a simple analog and digital precoding and combining technique for downlink mmWave CR systems exploiting short wavelength property of mmWave signal. The sum-rate performance of the proposed framework was analysed when the interference to primary user is not allowed to exceed beyond a certain limit. Finally, simulation results have been shown to exhibit the sum-rate performance under various conditions of the proposed framework. As a scope for future work, it would be interesting to develop the power allocation scheme for spatially correlated MIMO channel in mmWave band sharing systems.